\documentclass[prb,twocolumn,amsmath,amsfonts,longbibliography]{revtex4-2}
\usepackage{amssymb}
\usepackage{graphicx}% Include figure files
\usepackage{epsf}% Include figure files
\usepackage{epstopdf}
\usepackage{graphicx}
\usepackage{tikz}
\usepackage{float}
\usepackage{amsmath}
\usepackage{nccmath}
\usepackage[mathscr]{euscript}
\usepackage{natbib}
\usepackage{hyperref}

\begin{document}

\title{Detection of persistent current correlation in cavity-QED}
\author{Bogdan R. Bu{\l}ka}
\affiliation{Institute of Molecular Physics, Polish Academy of
Sciences, ul. M. Smoluchowskiego 17, 60-179 Pozna{\'n}, Poland}

\begin{abstract}

 We simulated the radiative response of  a cavity quantum electrodynamics (QED)  inductively coupled to the ring pierced by magnetic flux and analyzed its spectral dependence to get insight into persistent current dynamics.
Current fluctuations in the ring induce changes in the microwave resonator: shifting the resonant frequency and changing its damping. We use the linear response theory and calculate the current response function by means of the Green function technique.
Our model contains two quantum dots which divide the ring into two arms with different electron transfers.
There are two opposite (symmetric and asymmetric) components of the persistent current, which interplay can be observed the response functions.
The resonator reflectance shows characteristic shifts in the dispersive regime and avoided crossings at the resonance points.
The magnitude of the resonator frequency shift is greater for coupling to the arm with  higher transparency.
 Fluctuations of the symmetric component of the persistent current are relevant for a wide range of the Aharovov-Bohm phase $\phi$, while the asymmetric component becomes dominant close to $\phi\approx \pi$ (when the total persistent current changes its orientation).

\end{abstract}

\maketitle

%% \linenumbers

%% main text
\section{Introduction}

Recent technological developments in circuit quantum electrodynamics (cQED) offers efficient microwave resonators formed by superconducting Josephson junctions. This technique has been successfully applied to studies charge, spin and current dynamics in various nanodevices: single spins in doped crystals, quantum dot systems, superconducting qubits, nanomechanical oscillators or magnonic nanostructures \cite{Cottet2017,Burkard2020,Clerk2020,Blais2021}.
One could get insight into exotic condensed matter states, such as the
Kondo resonance and Majorana bound states \cite{Cottet2017,Burkard2020}, or to perform coherent manipulation of the Andreev states in superconducting qubits \cite{Janvier2015}.

Here, we want to show how cQED can be applied to measure correlations of the persistent current in a metallic ring pierced by magnetic flux. If  the ring size is small, smaller than the phase coherence length of electron waves ($L \ll L_{\phi}$), then quantum interference play a crucial role and manifest itself in electron transport. The persistent current was studied in many papers, both theoretically and experimentally (see the review \cite{Bleszynski2009} and references therein). However, an issue of fluctuations of the persistent current was undertaken only in several theoretical papers \cite{Cedraschi2000,Cedraschi2001,Moskalets2010,Semenov2010,Semenov2011,Komnik2014}.
In this paper we want to simulate the radiative response of the superconducting microwave resonator inductively coupled to the
ring with two quantum dots (2QD) connected by two different junctions, to get insight into quantum interference effects and persistent current dynamics.

\section{2QD ring and persistent current}

\begin{figure}\center
\includegraphics[width=.7\linewidth,clip]{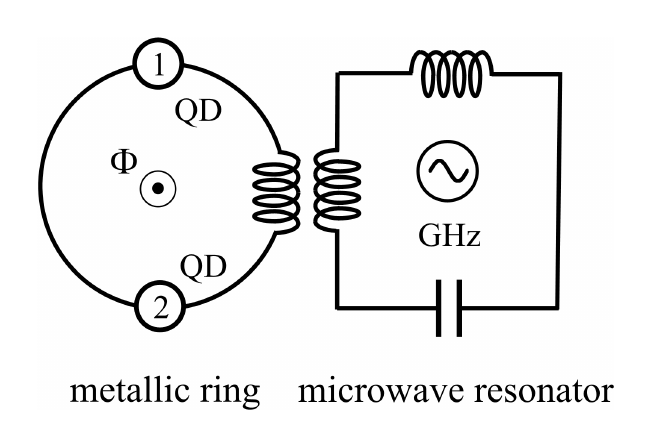}
\caption{Scheme of the considered system:
the ring of two quantum dots pierced by the magnetic flux $\Phi$, which is inductively coupled to the microwave resonator.
}\label{fig1}
\end{figure}

Fig.~\ref{fig1} presents our model of the 2QD ring pierced by magnetic flux.
Electrons in the ring are described by the Hamiltonian
\begin{align}
\label{eq:ham}
\hat{H} =   \sum_{\substack{i=1,2\\ \sigma=\uparrow,\downarrow}} \varepsilon_{i}c^{\dag}_{i\sigma} c_{i\sigma}
&+ \sum_{\sigma=\uparrow,\downarrow}(t_{12}\;c^{\dag}_{2\sigma} c_{1\sigma}+t_{21}\;c^{\dag}_{1\sigma} c_{2\sigma}) \; ,
\end{align}
where the first term corresponds the quantum dots with the single-level energy, $\varepsilon_{i}$. The second term is related with electron hopping between the dots: $t_{12}=t_L e^{i \phi/2}+t_R e^{-i \phi/2}$, $t_{21}=t_L e^{-i \phi/2}+t_R e^{i \phi/2}$, $t_L$ and $t_R$ is the hopping through the left (L) and right (R) arm of the ring. The hopping parameters include the phase shift $\phi =2\pi \Phi/\Phi_0$, due to presence of the magnetic flux $\Phi$, where $\Phi_0=\hbar/e$ denotes the one-electron flux quantum.

The persistent current operator is given by
\begin{align}
\hat{I}_{p}&=\frac{e}{\hbar}\frac{\partial \hat{H}}{\partial \phi}=(\hat{I}_{L}+\hat{I}_{R})/2,
\end{align}where
 \begin{align}
\hat{I}_L=\text{i}\frac{e}{\hbar}  t_L \sum_{\sigma}(e^{\text{i} \phi/2} c^{\dag}_{2\sigma} c_{1\sigma}-e^{-\text{i} \phi/2}c^{\dag}_{1\sigma} c_{2\sigma}),\label{eq:IL}\\
\hat{I}_R=\text{i}\frac{e}{\hbar}  t_R \sum_{\sigma}(e^{\text{i} \phi/2} c^{\dag}_{1\sigma} c_{2\sigma}-e^{-\text{i} \phi/2}c^{\dag}_{2\sigma} c_{1\sigma})\label{eq:IR}
\end{align}
are the current operators in the left and the right arm of the ring.
We use the Green function technique (details are in Appendix) to calculate the average currents, Eq.(\ref{eq:IL})-(\ref{eq:IR}),
\begin{align}
I_L=I_R=\frac{2e}{\hbar}\frac{ t_L t_R \sin(\phi)}{2\Delta} \left[f(E_+)-f(E_-)\right],
\end{align}
where $E_{\pm}=\epsilon \pm\Delta$ denote the energy levels, $\Delta=[\delta^2+t_L^2+t_R^2+2t_L t_R \cos(\phi)]^{1/2}$, $\epsilon=(\varepsilon_1+\varepsilon_2)/2$, $\delta=(\varepsilon_1-\varepsilon_2)/2$ and $f(\omega)$ is the Fermi factor.
The average currents are equal in both arms, but their transparencies are different, there are various competing local currents. To have insight into these processes we rewrite
$\Delta=[\delta^2+(t_L+t_R)^2\cos^2(\phi/2)+(t_L- t_R)^2 \sin^2(\phi/2)]^{1/2}$ and the persistent current
\begin{align}
I_p=&\frac{2e}{\hbar}\sum_{\nu=\pm}f(E_{\nu})\frac{\partial E_{\nu}}{\partial\phi}\nonumber\\
=&\frac{2e}{\hbar}\left[
\frac{(t_L+t_R)^2 \sin(\phi)}{8\Delta}-\frac{(t_L-t_R)^2 \sin(\phi)}{8\Delta}\right]\nonumber\\&\hskip 10em \times \left[f(E_+)-f(E_-)\right].
\end{align}
There are two opposite persistent currents related with the symmetric and asymmetric coupling. In the following sections, we will show that these competing processes can play an important role in current fluctuations.

\section{Current correlations}\label{corr}

\begin{figure}\center
\includegraphics[width=0.9\linewidth,clip]{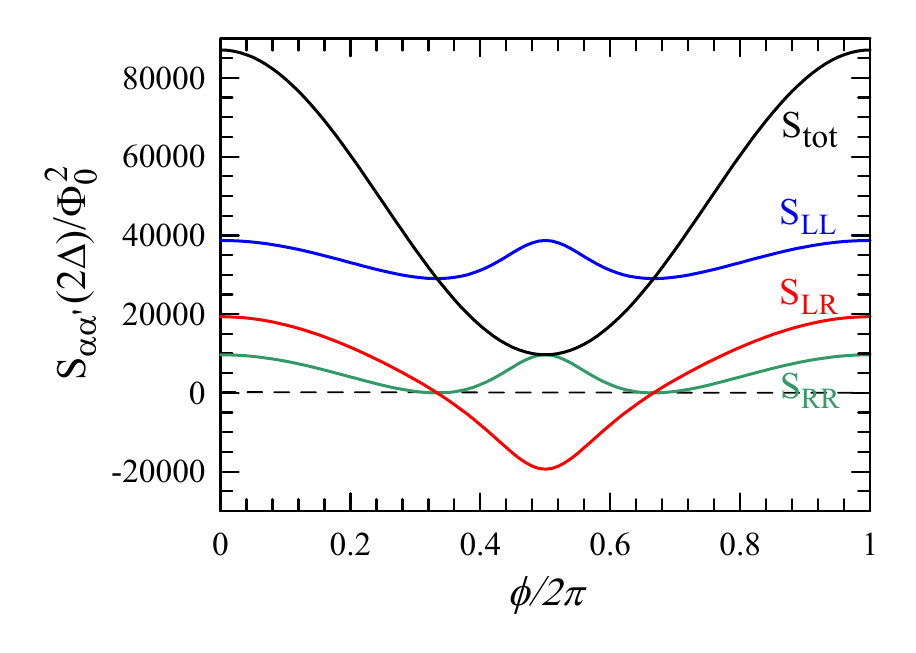}
\caption{Plot of the current correlation functions vs $\phi$: $S_{LL}(2\Delta)$ -blue, $S_{RR}(2\Delta)$ -green, $S_{LR}(2\Delta)$ -red and $S_{tot}(2\Delta)$ -black. }\label{fig2}
\end{figure}

The current-current correlation function (sometimes referred to as noise power) is defined as \cite{BlanterButtiker}
\begin{align}\label{noise}
  S_{\alpha,\alpha'}(\omega_p)&\equiv  \int d\tau \;e^{\text{i}\omega_p\tau}\langle \{ \delta \hat{I}_{\alpha}(\tau),\delta\hat{I}_{\alpha'}(0)\}\rangle,
\end{align}
where $\delta \hat{I}_{\alpha}(\tau)=\hat{I}_{\alpha}(\tau)- I_\alpha$ describes current fluctuation from its average value in the arm $\alpha=L,R$. The Green functions are used to determine $S_{\alpha,\alpha'}(\omega_p)$ --  see \ref{noisef}. To get insight into electron dynamics in the considered circuit we perform the spectral decomposition of the spectral density of the power noise [the integrant in (\ref{noise})]. One finds various relaxation processes but we take the  dominant ones in the limit of the weak coupling $\gamma$ to the thermal environment.

The zero-frequency correlator is given by
\begin{align}
S_{LL}(0)=\frac{2e^2}{\hbar}\frac{2 t_L^2 t_R^2 \sin^2(\phi)}{\gamma\Delta^2}F_0,
\end{align}
where $S_{LL}(0)=S_{RR}(0)=S_{LR}(0)=S_{RL}(0)$ -- compare with \cite{Moskalets2010}.
Here, we denoted the function $F_0= f(E_+)[1- f(E_+)]+f(E_-)[1- f(E_-)] $.

Persistent current fluctuations were considered by  Cedraschi and Buttiker \cite{Cedraschi2001} in a similar system, in a metallic ring with a single quantum dot. They showed that the dynamics of the ring can be considered as a  two-level system and its spectral densities exhibit two peaks.  In our 2QD ring the spectral densities show similar features, with two delta peaks at $\omega=\pm 2\Delta$ related with interlevel transitions. Therefore, for $\omega_p=2\Delta$ one can calculate the integral in (\ref{noise}) and express the correlation functions as
\begin{align}
S_{LL}(2\Delta)=&\frac{2e^2}{\hbar}\frac{ t_L^2 [\Delta^2 - t_R^2 \sin^2(\phi)]}{\gamma \Delta^2} F_{2\Delta},\\
S_{LR}(2\Delta)=&\frac{2e^2}{\hbar}\frac{t_L t_R [\Delta^2 \cos(\phi) + t_L t_R \sin^2(\phi)]}{\gamma \Delta^2} F_{2\Delta},\\
S_{RR}(2\Delta)=&\frac{2e^2}{\hbar}\frac{ t_R^2 [\Delta^2 - t_L^2 \sin^2(\phi)]}{\gamma \Delta^2} F_{2\Delta},\\
S_{tot}(2\Delta)=&\frac{2e^2}{\hbar}\frac{|t_{12}|^2}{\gamma}F_{2\Delta},
\end{align}
where  $F_{2\Delta}= f(E_+)[1- f(E_-)]+f(E_-)[1- f(E_+)]$.

Fig.\ref{fig2} presents these functions. The parameters of the 2QD ring are taken: $\epsilon=0$, $\delta=0$, $t_L/\hbar =2200$ MHz, $t_R/\hbar =1100$ MHz and $\gamma/\hbar=125$ MHz (close to the parameters of a double quantum dot capacitively coupled to a transmission line resonator \cite{Stockklauser2015}). Further in the paper, all quantities will be expressed in units of MHz to get a relation with an experiment. The total
current correlation (the black curve) monotonically decreases and reaches its minimum at $\phi=\pi$, when the persistent current changes its sign. The auto-correlation functions, $S_{LL}(2\Delta)$ and $S_{RR}(2\Delta)$, are non-monotonic functions with two minima at $\phi_{\pm}=\pi\pm\phi_0$, where $\phi_0=2 \arctan(\sqrt{|t_L+t_R|/|t_L-t_R|}$ for the considered case with $\delta=0$.
 The cross correlation function $S_{LR}(2\Delta)$ changes its sign exactly at the same values $\phi_{\pm}$.
It means that two opposite components of the persistent current compete with each other and their dynamics can be seen in the current correlation functions.  The asymmetric part of the persistent current dominates in the noise power in the range $\phi_-<\phi<\phi_+$, while the symmetric part is relevant outside this range.
 The next section is devoted to the detection of these processes.

\section{Response function}

\subsection{Derivation of the response function}

We assume that our 2QD ring is inductively coupled to  a superconducting microwave resonator (Fig.\ref{fig1}) and their interaction is described by the Hamiltonian
\begin{align}
\hat{H}_{int}= M\hat{I}_{\alpha}\hat{I_r},
\end{align}
where $M$ is the mutual inductance, $\hat{I}_{\alpha}$ denotes the current operator in the considered 2QD ring. The current operator in the resonator can be expressed as $\hat{I_r}=\omega_r\sqrt{\hbar/2Z}(\hat{a}^\dag+\hat{a})$ \cite{Omelyanchouk2010,Janvier2015,Forn-Diaz2019}, where $\omega_r$ denotes its frequency, $Z$ - the impedance and $\hat{a}^\dag$ ($\hat{a}$) is the creation (annihilation) photon operator.
 The resonator is described by a simple harmonic oscillator with the Hamiltonian $\hat{H}_{r}= \hbar\omega_r(\hat{a}^\dag \hat{a}+1/2)$.

In an experiment a continuous microwave drive is applied to the input port of the resonator to control and measure the photon system ~\cite{Blais2021,Stockklauser2015}. To describe perturbation of the resonator signal by the attached nano-system the semiclassical input-output theory~\cite{Walls2008} is used (for a linear coupling between both systems and  confining to single photon processes)  ~\cite{Dmytruk2016,Blais2021,Kratochwil2021}.
For the single sided configuration the reflection coefficient can be expressed as
\begin{align}\label{eq:s11}
S_{11} \equiv\frac{a_{out}}{a_{in}}
=-\frac{\omega_p - \omega_r + \text{i} (\kappa_{i} - \kappa_{e})/2- g^2\chi^r(\omega_{p})}{\omega_p - \omega_r  + \text{i}(\kappa_{i} +\kappa_{e})/2-g^2\chi^r(\omega_{p})},
\end{align}
where  $\omega_r-\omega_p$ is the detuning of the resonator frequency from the probe (drive) frequency $\omega_p$, $\kappa_{i}$ and $\kappa_{e}$ denote internal and external resonator dissipation rates.
$\chi^r$ denotes the response function and $g$ is a coupling of the nanosystem with the resonator. In an experiment the complex reflection coefficient $S_{11}=|S_{11}|e^{i\varphi}=I +i Q$ is determined, with its amplitude $|S_{11}|$ as well as the phase $\varphi$ or equivalently the field quadratures $I$ and $Q$.

The current-current response function (the current susceptibility) is derived within the linear response theory as
\begin{align}\label{eq:chiR}
\chi^{r}_{\alpha\alpha'}(\omega_p)\equiv -\frac{\text{i}}{\hbar}\int d\tau\; e^{\text{i}\omega_p\tau}\theta(\tau) \langle[\hat{I}_{\alpha}(\tau),\hat{I}_{\alpha'}(0)]\rangle_{g=0},
\end{align}
which is calculated for the isolated 2QD ring, described by the Hamiltonian (\ref{eq:ham}).
This is the key quantity of interest in measurement of the reflection coefficient.

\subsection{Response of the 2QD ring}

\begin{figure}\center
%\centering
\includegraphics[width=1\linewidth,clip]{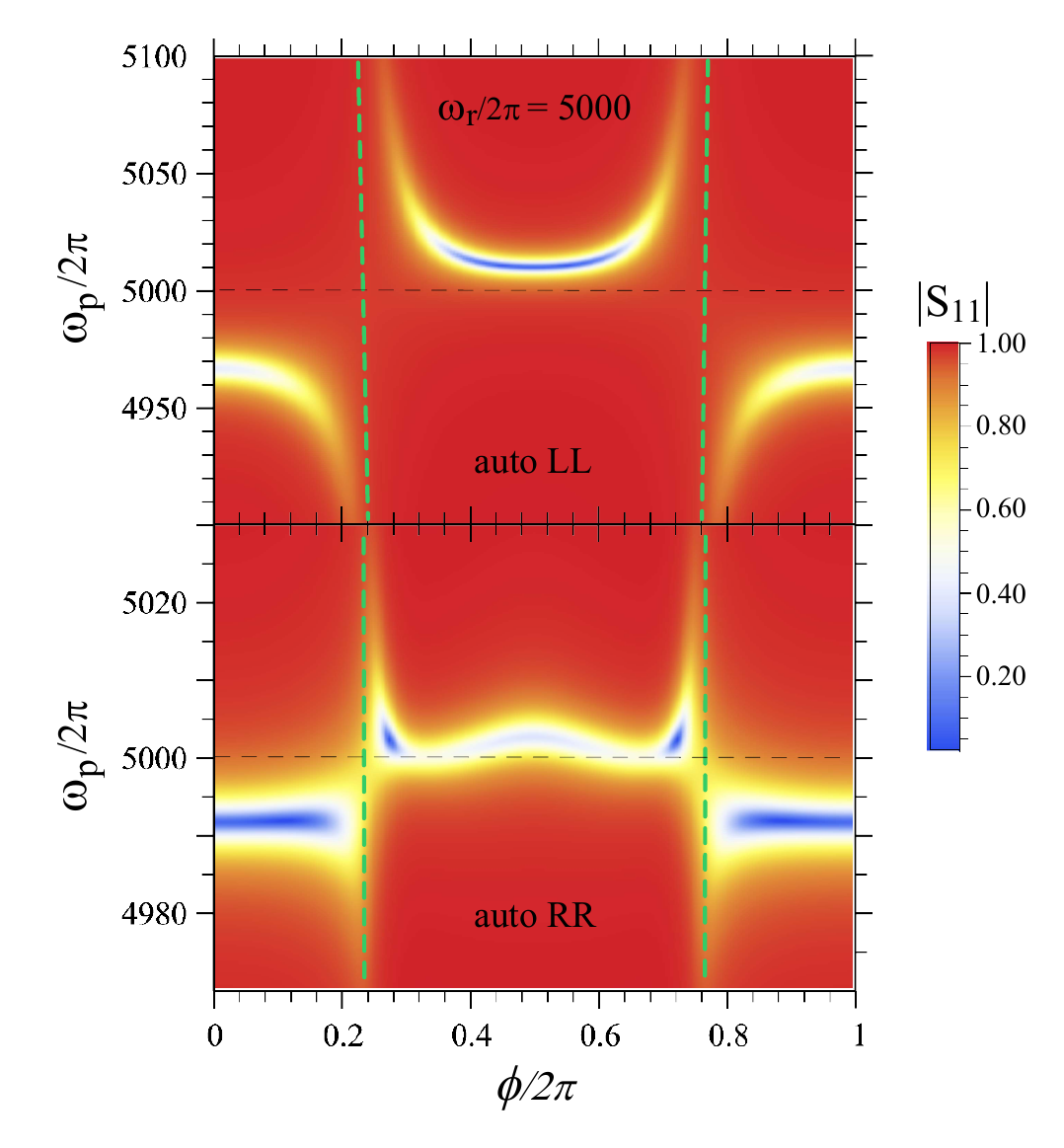}
\caption{ Resonator reflectance $|S_{11}|$ as a function of $(\phi, \omega_p)$ for coupling to the left and the right arm (with the hopping parameters: $t_L/\hbar =2200$ MHz, $t_R/\hbar =1100$ MHz) at the resonator frequency $\omega_r/2\pi=5000$ MHz.  The green dashed curve presents the dispersion relation of the 2QD ring: $\hbar\omega_p = 2\Delta$. The other parameters of the 2QD ring are the same as in Fig.\ref{fig2}
}
\label{plot5000}
\end{figure}

The response function $\chi^{r}_{\alpha,\alpha'}(\omega_p)$ for our 2QD ring is calculated by means of the Green function technique -- see \ref{responsef}.  Next, we calculate the resonator reflectance $|S_{11}|$ as a map with respect of $(\phi, \omega_p)$ -- the result is exhibited in Fig.~\ref{plot5000} for the left and the right arm of the ring (the upper and the bottom plot, respectively).
We take the resonator parameters $\kappa_i/2\pi=$ 1 MHz, and  $\kappa_e/2\pi=$ 4 MHz (as in the experiment \cite{Hays2021}) and the coupling $g^2=\lambda^2\Phi_0^2$ with $\lambda=0.05$ \cite{Metzger2021,Hermansen2022}. For  the considered 2QD ring the  excitation energy is in the range: 2200 MHz $< 2\Delta/\hbar <$ 6600 MHz.
We assume the resonator frequency $\omega_r/2\pi=$ 5000 MHZ, which  crosses the excitation spectrum of the 2QD ring (shown as the green dashed curve in the figure). Fig.~\ref{plot5000} shows characteristic shifts in the dispersive regime and
 avoided crossings at the resonance points.
Notice that response of the left and the right arm are different, due to their different transparencies (large and small in the L and R arm, respectively) -- compare with the current correlations $S_{LL}$ and $S_{RR}$ in Fig.~\ref{fig2}.

\begin{figure}\center
%\centering
\includegraphics[width=1.0\linewidth,clip]{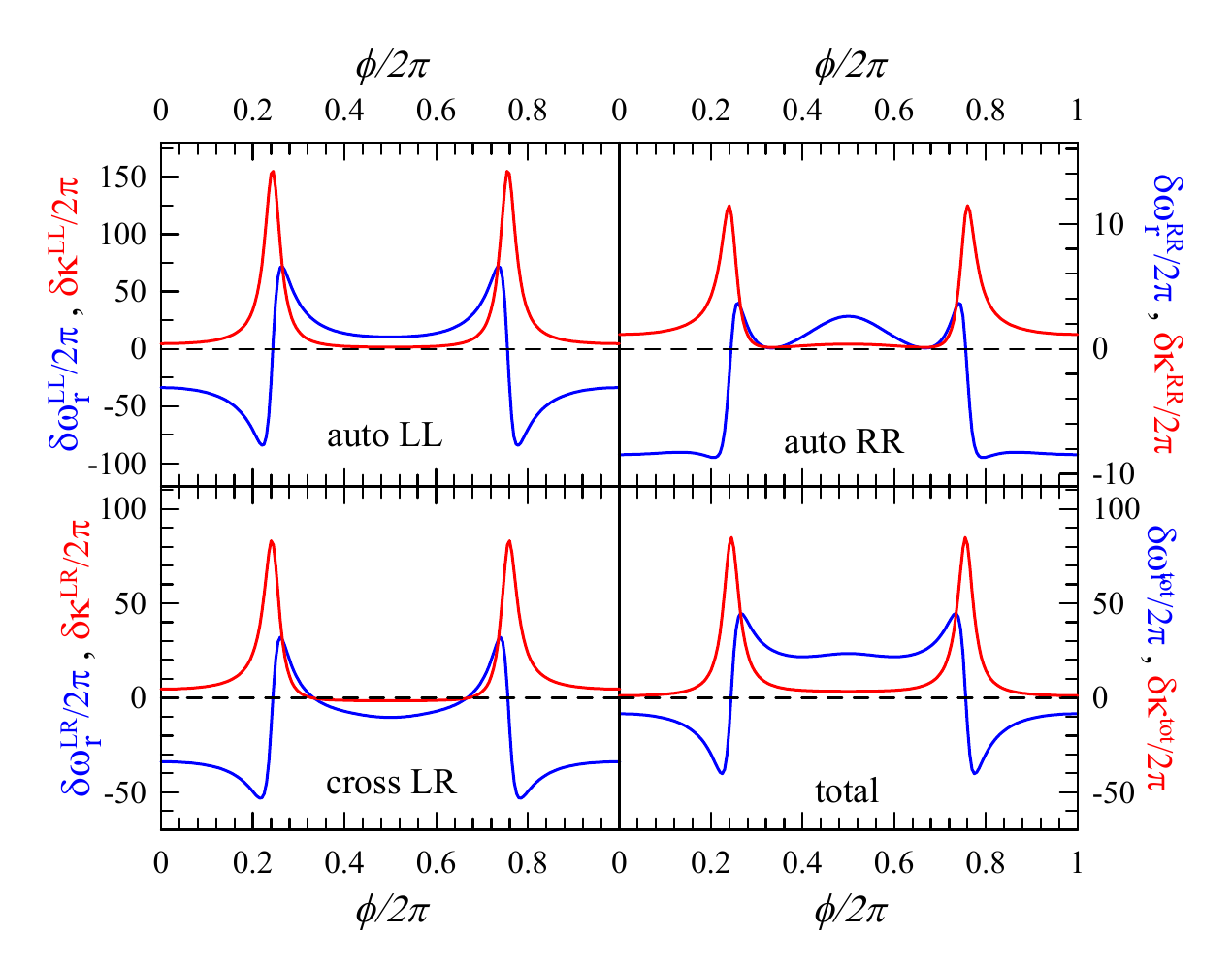}
\caption{Resonator frequency shift $\delta\omega_r^{\alpha\alpha'}/2\pi$ (blue curve)  and its broadening $\delta\kappa^{\alpha\alpha'}/2\pi$ (red curve) for the various response configurations ($\alpha, \alpha'= L,R$) plotted as a function of $\phi$  at the resonant frequency $\omega_r/2\pi$ = 5000 MHz. }
\label{shift5000}
\end{figure}

From the reflection spectra, Eq.(\ref{eq:s11}), one can also extract the frequency shift $\delta\omega_r^{\alpha\alpha'}$ and the damping ratio $\delta\kappa^{\alpha\alpha'}$ of the resonator. They are related with the real and the imaginary part of the response function
\begin{align}
\delta\omega_r^{\alpha\alpha'}&=(g^2/\hbar)\Re[\chi^r_{\alpha\alpha'}(\omega_{r})],\\
\delta\kappa^{\alpha\alpha'}&=-(g^2/\hbar)\Im[\chi^r_{\alpha\alpha'}(\omega_{r})].
\end{align}
Fig.~\ref{shift5000} shows these quantities calculated for various components of the response function.
Close to the resonance one can see a characteristic Lorentzian shape of the response function,   with  $\delta\omega_r^{\alpha\alpha'}=0$ and a maximum for $\delta\kappa^{\alpha\alpha'}$.
The amount of the frequency shift $\delta\omega_r^{LL}$ and $\delta\omega_r^{RR}$ is different (an order of magnitude smaller in the R arm). It is seen that $\delta\omega_r^{RR}$ is a non-monotonic function and reaches its minimum at $\phi_{\pm}=\pi\pm\phi_0$.
This is manifestation of interplay of two opposite persistent currents in the 2QD system. We also present the plot for $\delta\omega_r^{LR}$, corresponding to the cross current correlations. This is an interesting result, showing a negative frequency shift in the region $\phi_-<\phi<\phi_+$ where the asymmetric component of the persistent current dominates. Unfortunately, a direct measurement of the cross current response is impossible. One can measure the response for the total current in a circuit with a double coupling to both arms of the ring. Using $\delta\omega_r^{LL}$ and $\delta\omega_r^{RR}$ (determined in earlier measurements) one can extract the cross correlation component.

\section{Summary }

We have shown that two opposite persistent currents are present in the asymmetric ring and their interplay can be detected by measurement the radiative response of the inductively coupled microwave resonator in cQED.
The symmetric component dominates in a wide range of $\phi$, while the asymmetric component is relevant close to $\phi\approx \pi$ and its range increases with the asymmetry factor. Their fluctuations play different roles, which can be seen in the response functions, in the frequency shift of the resonator $\delta\omega_r^{\alpha\alpha'}$ and its damping factor $\delta\kappa^{\alpha\alpha'}$.

Our results are exact in the single electron case, however, one can extend the studies for many electrons and take into account Coulomb interactions. In such the case the ac linear current response should be calculated in the Lehmann representation, following the Kubo formalism described  in Ref.\cite{Trivedi1988}. Calculating eigenvalues of the system one should keep in mind that the charge and the phase are quantum conjugate variables.~\cite{Cedraschi2000,Cedraschi2001}.

The Aharonov-Bohm effect in an 2QD system was demonstrated experimentally in \cite{Holleitner2001}. Such the system one could capacitively coupled to a microwave resonator (as in \cite{Stockklauser2015}) and perform studies of charge dynamics.  Inductive coupling of the 2QD ring to the resonator can be an experimental challenge. Recent achievements in superconducting quantum circuits show that galvanic inductive couplings can reach an ultra-strong regime, in particular for flux qubits \cite{Forn-Diaz2019,Blais2021}. We believe that  such the technique can be also applied to studies persistent current dynamics the considered 2QD ring.

It is worth to mention that the considered model in many aspects is similar to the model of the Josephson junction with a quantum dot inside \cite{Martin-Rodero2011}. There, one can expect also interplay of two opposite Josephson currents which can be detected by cQED.

\appendix

\section{Green function method}
\setcounter{equation}{0}
\renewcommand\theequation{A.\arabic{equation}}

To calculate the average current and the current-current correlation functions we use the lesser and the greater Green function: $\hat{G}^{<}(\omega)= f(\omega)[\hat{G}^{a}(\omega)-\hat{G}^{r}(\omega)]$ and $\hat{G}^{>}(\omega)= [f(\omega)-1][\hat{G}^{a}(\omega)-\hat{G}^{r}(\omega)]$ , where $f(\omega)$ denotes  is the Fermi factor, while
the retarded and advanced Green functions  are expressed as
\begin{gather}
\hat{G}^{r,a}(\omega)=  \left[\begin{array}{cc}
  \hbar\omega\pm \text{i}\gamma -\varepsilon_1 & t_{21}\\
 t_{12}& \hbar\omega\pm \text{i}\gamma -\varepsilon_2
  \end{array}\right]^{-1}.
\end{gather}
Here, a small parameter $\gamma$ is introduced to take into account thermal dissipation processes in the system.

\subsection{Calculation of power noise}\label{noisef}

The current-current correlation function $S_{\alpha\alpha'}$, Eq.(\ref{noise}), can be written as
\begin{align}
S_{\alpha\alpha'}=&(4e^2/\hbar)[t_{12}^{\alpha}t_{12}^{\alpha'} a_{12,12}-
t_{12}^{\alpha}t_{21}^{\alpha'}  a_{12,21}\nonumber\\ &-t_{21}^{\alpha}t_{12}^{\alpha'}  a_{21,12}
+t_{21}^{\alpha}t_{21}^{\alpha'}  a_{21,21}],
\end{align}
where $t_{12}^L=t_L e^{i \phi/2}$, $t_{12}^R=t_R e^{-i \phi/2}$,  $t_{21}^L=t_L e^{-i \phi/2}$, $t_{21}^R=t_R e^{-i \phi/2}$ and $t_{12}=t_{12}^L+t_{12}^R$, $t_{21}=t_{21}^L+t_{21}^R$.
The coefficients are expressed by two particle averages; for example
\begin{align}
a_{12,12}=\int d\omega\; e^{\imath \omega_p t}&[\langle c^{\dag}_{1}(t)c_2(t)c^{\dag}_1(0)c_2(0)\rangle \nonumber\\&-\langle c^{\dag}_{1}(t)c_2(t)\rangle \langle c^{\dag}_1(0)c_2(0)\rangle].
\end{align}
Next, these averages are decoupled by means of Wick’s theorem to products of single particle averages.
Using the Green functions the coefficients are expressed as
\begin{align}
a_{12,12}= \int \frac{d\omega}{2\pi}[G^<_{2,1}(\omega) G^>_{2,1}(\omega_+) +G^<_{2,1}(\omega_+) G^>_{2,1}(\omega)],\nonumber\\
a_{21,21}= \int \frac{d\omega}{2\pi}
   [G^<_{1,2}(\omega) G^>_{1,2}(\omega_+) +G^<_{1,2}(\omega_+) G^>_{1,2}(\omega)],\nonumber\\
a_{12,21}= \int \frac{d\omega}{2\pi} [G^<_{1,1}(\omega) G^>_{2,2}(\omega_+) +G^<_{1,1}(\omega_+) G^>_{2,2}(\omega)],\nonumber\\
a_{21,12}=  \int \frac{d\omega}{2\pi}   [G^<_{2,2}(\omega) G^>_{1,1}(\omega_+) +G^<_{2,2}(\omega_+) G^>_{1,1}(\omega)],
\end{align}
where $G^{<(>)}_{i,j}$ denote the elements of the lesser (greeter) Green function and $\omega_{+}=\omega+\omega_p$.

\subsection{Calculation of response function}\label{responsef}

The current susceptibility $\chi^{r}_{\alpha\alpha'}$, Eq.(\ref{eq:chiR}),
is also determined by means of the Green function technique, following
Ref. \cite{Bruhat2016,Cottet2020}. In principle, this function can be calculated by other techniques (e.g. see \cite{Hays2021,Metzger2021,Hermansen2022}).
Its Fourier transform can be expressed as
\begin{align}\label{eq:chir}
\chi^{r*}_{\alpha,\alpha'}=&(4e^2/\hbar^2)[t_{12}^{\alpha}t_{12}^{\alpha'} b_{12,12}-
t_{12}^{\alpha}t_{21}^{\alpha'}  b_{12,21}\nonumber\\ &-t_{21}^{\alpha}t_{12}^{\alpha'}  b_{21,12}
+t_{21}^{\alpha}t_{21}^{\alpha'}  b_{21,21}],
\end{align}
where
\begin{align}
b_{12,12}=& -\text{i}\int \frac{d\omega}{2\pi} G^<_{12}(\omega)[ G^r_{12}(\omega_+) + G^a_{12}(\omega_-)],\nonumber\\
b_{12,21}=& -\text{i}\int \frac{d\omega}{2\pi}[G^<_{11}(\omega) G^r_{22}(\omega_+) +G^<_{22}(\omega) G^a_{11}(\omega_-)],\nonumber\\
b_{21,12}=& -\text{i}\int \frac{d\omega}{2\pi}[G^<_{22}(\omega) G^r_{11}(\omega_+) +G^<_{11}(\omega) G^a_{22}(\omega_-)],\nonumber\\
b_{21,21}=&- \text{i}\int \frac{d\omega}{2\pi}G^<_{21}(\omega)[ G^r_{21}(\omega_+) + G^a_{21}(\omega_-)]
\end{align}
and $\omega_{\pm}=\omega\pm \omega_p$.

\end{document}